\documentclass[12pt]{article}
\usepackage{graphicx}

\begin{document}

\baselineskip=30pt

\begin{titlepage}

\vskip1in

\begin{center}
\LARGE{\bf N-loop Treatment of Overlapping Diagrams by the
Implicit Regularization Technique.}
\end{center}

\vskip1.0cm
\begin{center}

\Large{ S. R. Gobira$^{1,2}$ \ and  M. C. Nemes$^{2}$}\\ 
\end{center}

\vskip1.0cm
\begin{center}
$(1)$UFT - Universidade Federal do Tocantins\\
CEP-77020-210, Palmas - To - Brazil\\
\vskip1.0cm
$(2)$UFMG - Universidade Federal de Minas Gerais\\
Physics Department - ICEx\\
P.O. BOX 702, 30.161-970, Belo Horizonte - MG - Brazil\\
\vskip1.0cm

\vskip0.5cm

{\it {gobira@uft.edu.br and carolina@fisica.ufmg.br}}
\end{center}

\newpage
\begin{abstract}
\noindent
We show how the Implicit Regularization Technique (IRT) can be used for the 
perturbative renormalization of a simple  field theoretical model, generally
used as a test theory for new techniques. While IRT has been applied successfully in many problems
involving symmetry breaking anomalies and nonabelian gauge groups, all at one loop level, 
this is the first attempt to a generalization of the technique for perturbative  renormalization.
We show that the overlapping divergent loops can be given a completely algebraic treatment.
We display the connection between renormalization and
counterterms in the Lagrangian. The algebraic advantages make IRT worth studying for
perturbative renormalization of gauge theories.

\end{abstract}
\end{titlepage}

\section{\protect\bigskip Introduction}

Quantum field theoretical predictions of physical quantities should in
principle be independent of the particular scheme used to renormalize the
theory. The renormalization program allows to get rid of the singularities
by redefinition of the parameters in the Lagrangian in a consistent way for
a renormalizable model. Also, in this process we must make sure that the
relevant symmetries of the underlying theory are preserved and therefore
avoid the appearance of spurious anomalies which otherwise would have to be
controlled order by order in perturbation theory by imposing symmetry
restoring constraint equations.

As for the existing regularization schemes whilst for the theories with low
symmetry content nearly all regulators do a good job, this is not the case
for most theories of particle interactions in which gauge symmetry,
supersymmetry (SUSY) and so on play a fundamental role. Dimensional
Regularization (DR) \cite{2}\cite{3}\cite{4} is an efficient and pragmatical
method which explicitly preserves gauge symmetry. However in the presence of
dimension specific objects such as $\gamma ^{5}$ matrices, a suitable
generalization of the Dirac algebra must be constructed to be compatible
with analytical continuation on the space-time dimension. This is the case
of the Electroweak sector of the Standard Model. Since chiral symmetry is
broken in this case, the corresponding Ward-Slavnov-Taylor identities must
be imposed order by order, what turns the computations beyond one loop order
very hard.

For SUSY theories, the fact that the equality between Bose and Fermi degrees
of freedom only holds for specific values of the space time dimension, SUSY
is broken in DR. A naive scheme (Dimensional Reduction) in which the field
components are left unchanged while the loop integrals are performed in d
dimensions can be shown to be inconsistent, see reference.\cite{5}. Similar
problems arise in Chern-Simons field theories in which the Levi-Civitta
tensor is the three dimensional analogous of the $\gamma ^{5}$ matrix \cite
{6}\cite{7}.

A particularly interesting regularization independent framework is the
Differential Renormalization program pioneered by Freedman et al \cite{8}.
The basic idea of this scheme is that renormalization comes from the fact
that products of propagators must be extended to be distributions so that a
Fourier transform is well defined. Working in (Euclidean) coordinate space
one writes the amplitude as a derivative of a distribution less divergent at
coincident points. The derivatives are understood in the sense of
distribution theory, i.e. acting formally by parts. The amplitudes written
in this way are identical to the bare ones for separate points but behave
well at coincident points. An intrinsic arbitrary scale appears in this
process which is used as a Callan-Symanzik renormalization group parameter.
The advantage of this method is that it works in integer space-time
dimension, and it has been shown to yield satisfactory results where it was
tested \cite{9}\cite{10}\cite{11}\cite{12}. However no general procedure
using Differential Renormalization beyond one loop order, such that gauge
invariance is automatic, has been constructed yet.

Recently an essentially regularization independent procedure has been
advanced (see refs.\cite{13}\cite{14}\cite{15}\cite{16}\cite{ijtp}). It
presents the same consistency as Differential Regularization, working
however in momentum space. It has been recently shown that the method
respects both abelian and nonabelian symmetries at a few loops level and the
infrared divergence can be treated by the same procedure constructed for
ultraviolet divergences, without any modification, as has been shown for the
gluon self energy \cite{16}. All these results indicate that the method
deserves further investigation, for example, the question of n-loops and
perturbative renormalization in a theory. Encouraged by the previous results
and also aware of the difficulties of \ setting up a consistent scheme for
perturbative renormalization where there are already so many of the, so well
established, we feel for the sake of completeness of the Implicit
Regularization Technique to take one step in this direction This is
precisely the purpose of present contribution,Since this is a major task,
already pursued by several sophisticated schemes, we should like to start by
investigating a simple, although non trivial theory (involving, e.g.
overlapping divergences) to test our method further. If we succeed, it will
be most important first step to construct a perturbatively n-loop
renormalization scheme in the spirit and consistency of e.g. BPHZ and
others. For our so far very humble purposes, we study specifically the $\phi
_{6}^{3}$ theory.

The IRT is essentially regularization independent in the sense that a
specific regulator needs never be used. A convenient identity at level of
the integrand enables us to rewrite the amplitude as a sum of three types of
contributions namely local divergences (basic divergent integrals which
characterize the divergent structure of theory), nonlocal divergences
typical of divergent sub-structures contributions) and finite contributions.
The local divergences obtained in this way are equivalent to those obtained
by performing a Taylor expansion (like in BPHZ) only for primitive diagrams,
with no subdivergences. In the case subdivergences are present some more
subtle differences between our method and BPHZ should be pointed out: a)
Taylor expansions modify the original Feynman integrand; this procedure may
therefore violate symmetries. We circumvent this problem by means of using a
mathematical identity, preserving thus the original content of the original
Feynman amplitude. b) In order to classify the subdivergences, BPHZ uses
graphic representation (forest formula). In our procedure we identify, in
terms of integrals, divergences which occurred at lower loop orders. This
avoids in particular the complicated topological graphs structures and
substitutes this step by an algebraic procedure, (at least in this simple
case)from which the counterterms appear in a natural systematic way.
Specific examples are given in the text. Moreover, just like Differential
Renormalization arbitrary local terms can be duly parametrized and properly
adjusted on physical grounds. This is particularly important for finite
renormalization in order to clear the calculation from regularization
ambiguities. Finally our framework lives in the integer space-time dimension
which avoids well-known problems with dimension specific theories.

In order to illustrate our method we study the renormalization of $\phi
_{6}^{3}$ theory to n-loop order.

\section{Renormalization by the Implicit Regularization Technique}

In this section we construct an extension of a technique firstly designed
for one\cite{13}\cite{14} and two loop calculations \cite{15} for performing
a $n^{th}$order renormalizability proof.

In order to illustrate the procedure, consider the following divergent
amplitude, typical of one loop order: 
\begin{equation}
\int_{\Lambda }\frac{d^{4}k}{(2\pi )^{4}}\frac{1}{%
[(k+p)^{2}-m^{2}](k^{2}-m^{2})}\cdot  \label{1}
\end{equation}
The symbol $\Lambda $ under the integral sign presupposes, as discussed, an
implicit regularization. Now, in order to separate the logarithmic
divergence from the finite part, we use the following identity in the factor
involving the external momentum $p$: 
\begin{eqnarray}
\frac{1}{[(k+p)^{2}-m^{2}]} &=&\sum_{j=0}^{N}\frac{\left( -1\right)
^{j}\left( p^{2}+2p\cdot k\right) ^{j}}{\left( k^{2}-m^{2}\right) ^{j+1}} 
\nonumber \\
&&+\frac{\left( -1\right) ^{N+1}\left( p^{2}+2p\cdot k\right) ^{N+1}}{\left(
k^{2}-m^{2}\right) ^{N+1}[\left( k+p\right) ^{2}-m^{2}]}\cdot  \label{2}
\end{eqnarray}
In the above expression $N$ is chosen so that the last term is finite under
integration over $k$. Notice also that in the first term in equation (\ref{2}%
), the external momentum appears only in the numerator and thus after
integration it can yield at most polynomials in $p$ multiplied by
divergences. For our present example we need $N=0$, since we are dealing
with a logarithmic divergence. We can rewrite (\ref{1}) using (\ref{2}) as 
\begin{equation}
I=\int_{\Lambda }\frac{d^{4}k}{(2\pi )^{4}}\frac{1}{(k^{2}-m^{2})^{2}}-\int 
\frac{d^{4}k}{(2\pi )^{4}}\frac{p^{2}+2p\cdot k}{%
[(k+p)^{2}-m^{2}](k^{2}-m^{2})^{2}}\,\cdot  \label{3}
\end{equation}
Now only the first of these two integrals is divergent. The others can be
easily integrated out to yield 
\begin{equation}
I=I_{log}(m^{2})-\frac{i}{(4\pi )^{2}}Z_{0}(m^{2},p^{2})  \label{4}
\end{equation}
where 
\begin{equation}
I_{log}(m^{2})=\int_{\Lambda }\frac{d^{4}k}{(2\pi )^{4}}\frac{1}{%
(k^{2}-m^{2})^{2}}  \label{5}
\end{equation}
and 
\begin{equation}
Z_{0}(m^{2},p^{2})=\int_{0}^{1}dz\,\ln \Big(\frac{p^{2}z(1-z)-m^{2}}{-m^{2}}%
\Big).\cdot  \label{6}
\end{equation}
Note that, since no explicit form for the regulator has been used, one can
make immediate contact with other regularizations. Details of calculations
of several one loop amplitudes and their associated Ward identities by using
this method can be found in \cite{14}.

By convenience we divide the diagrams which contribute to a given order in
two classes: the first which does not contain diagrams which possess two
point functions as subdivergences and in the second class those which do.

Let us start with the first class of diagrams. To show how the procedure
works it is enough to consider a general Feynman amplitude with one external
momentum $p$, one coupling constant $\lambda $ and one mass parameter $m$.
We work in the $4$-dimensional space-time although the generalization to any
integer dimension is straightforward. We denote by $q$ a sum of internal
momenta $k_{i}$. The amplitude in question can always be written as 
\begin{equation}
\Gamma =\prod_{i=1}^{n}\int_{\Lambda }\frac{d^{4}k_{i}}{(2\pi )^{4}}%
R(p,q,m,\lambda )\left[ \prod_{j=1}^{l}f_{j}(p,q_{j},m^{2})\right]  \label{7}
\end{equation}
where 
\begin{equation}
f_{j}(p,q_{j},m^{2})=\frac{1}{[(p-q_{j})^{2}-m^{2}]}  \label{8}
\end{equation}
and 
\[
\mbox{l}=\mbox{number of}\,\,\ f\,\,\,\mbox{structures} 
\]
\[
\mbox{n}=\mbox{number of loops.}\cdot 
\]

Note that we have explicitly separated the terms involving the external
momentum in the denominator, from which nonlocal divergent contributions can
arise after integration over the internal momenta. The structure $%
R(p,q,m,\lambda )$ contains all other ingredients of the amplitude such as
coupling constants, results of Dirac traces, and so on.

For simplicity we adopt the following notation 
\begin{equation}
\Gamma =(\Pi R)(\Pi f)  \label{9}
\end{equation}
where 
\begin{equation}
(\Pi R)=\prod_{i=1}^{n}\int_{\Lambda }\frac{d^{4}k_{i}}{(2\pi )^{4}}%
R(p,q,m,\lambda )  \label{10}
\end{equation}
and 
\begin{equation}
(\Pi f)=\prod_{j=1}^{l}f_{j}(p,q_{j},m^{2})\cdot  \label{11}
\end{equation}
As discussed before the source of all possible troubles in the
renormalization process will arise from the structure $(\Pi f).$Our method
focus attention on these structures. In order to clearly separate finite,
local divergences (whose dependence on the external momenta is only a
polynomial) from the nonlocal divergences we use a strategy which is
completely based on the identity (\ref{2})

Define the operator $T^{D}$ which acts on {\it each} structure $f$ in the
following way 
\begin{equation}
T^{0}f=\frac{1}{q_{j}^{2}-m^{2}}+\frac{2p.q_{j}-p^{2}}{(q_{j}^{2}-m^{2})}%
\left\{ \frac{1}{[(p-q_{j})^{2}-m^{2}]}\right\}  \label{12}
\end{equation}
\begin{equation}
T^{1}f=\frac{1}{q_{j}^{2}-m^{2}}+\frac{(2p.q_{j}-p^{2})}{%
(q_{j}^{2}-m^{2})^{2}}+\frac{(2p.q_{j}-p^{2})^{2}}{(q_{j}^{2}-m^{2})^{2}}%
\left\{ \frac{1}{[(p-q_{j})^{2}-m^{2}]}\right\}
\end{equation}
\begin{eqnarray}
T^{2}f &=&\frac{1}{q_{j}^{2}-m^{2}}+\frac{(2p.q_{j}-p^{2})}{%
(q_{j}^{2}-m^{2})^{2}}+\frac{(2p.q_{j}-p^{2})^{2}}{(q_{j}^{2}-m^{2})^{3}} 
\nonumber \\
&&+\frac{(2p.q_{j}-p^{2})^{3}}{(q_{j}^{2}-m^{2})^{3}}\left\{ \frac{1}{%
[(p-q_{j})^{2}-m^{2}]}\right\} \cdot  \label{14}
\end{eqnarray}
Note that the action of the operator $T^{D}$ is equivalent to a Taylor
expansion around zero external momentum where the first terms are kept and
the rest of the series is resumed, yielding thus a convenient identity. Note
also that the degree of divergence of the various terms is decreasing.

The procedure we have in mind consists of applying the operation, in a
particular amplitude with superficial degree of divergence $D$, to {\it each}
function $f_{j}$%
\begin{equation}
T^{D}\Gamma =(\Pi R)\prod_{j=1}^{l}T_{j}^{D}f_{j}(p,q_{j},m^{2})\cdot
\label{15}
\end{equation}
The result of the operation will always have the form 
\begin{equation}
T^{D}f(p,q,m^{2})=f^{div}(p,q,m^{2})+f^{fin}(p,q,m^{2})\cdot  \label{16}
\end{equation}
We define 
\begin{equation}
f^{div}(p,q,m^{2})=\sum_{i=0}^{D}f^{i}(p,q,m^{2})\cdot  \label{17}
\end{equation}
Let us exemplify. Take a quadratically divergent amplitude. To each
contribution of the form 
\[
\frac{1}{(p-q_{j})^{2}-m^{2}} 
\]
we associate

\begin{equation}
f^{0}(q,m^{2})=\frac{1}{q^{2}-m^{2}}  \label{18}
\end{equation}

\begin{equation}
f^{1}(p,q,m^{2})=\frac{2p.q-p^{2}}{(q^{2}-m^{2})^{2}}  \label{19}
\end{equation}
\begin{equation}
f^{2}(p,q,m^{2})=\frac{(2p.q)^{2}}{(q^{2}-m^{2})^{3}}  \label{20}
\end{equation}
and 
\begin{equation}
f^{fin}(p,q,m^{2})=\frac{p^{4}-4p^{2}(p.q)}{(q^{2}-m^{2})^{3}}+\frac{%
(2p.q-p^{2})^{3}}{(q^{2}-m^{2})^{3}[(p-q_{j})^{2}-m^{2}]}\cdot  \label{21}
\end{equation}
The definitions (\ref{18}),(\ref{19}),(\ref{20}),(\ref{21}) are not unique.
It is simply convenient for our purposes. Using these we rewrite the
amplitude as a sum of various contributions. According to our notation 
\begin{equation}
T^{D}\Gamma =(\Pi
R)\prod_{j=1}^{l}[f_{j}^{div}(p,q,m^{2})+f_{j}^{fin}(p,q,m^{2})]\cdot
\label{22}
\end{equation}
In this way we can identify three distinct contributions for the amplitude 
\begin{equation}
T^{D}\Gamma =\Gamma _{fin}^{1}+\Gamma _{local}+\Gamma _{nonlocal}  \label{23}
\end{equation}
where 
\begin{equation}
\Gamma _{fin}^{1}=(\Pi R)\prod_{j=1}^{l}f_{j}^{fin}(p,q,m^{2})\cdot
\label{24}
\end{equation}
The second contribution contains only local divergences and, for some
particular $(\Pi R)$ structures, it can contain finite contributions too. It
is identified as 
\begin{eqnarray}
\Gamma _{local} &=&(\Pi R)\prod_{j=1}^{l}f_{j}^{div}(p,q,m^{2})  \nonumber \\
&=&\Gamma _{fin}^{2}+\Gamma _{local}^{div}\cdot  \label{25}
\end{eqnarray}
These local divergences correspond to counterterms which are characteristic
of the order we are renormalizing. For example, they can have the form 
\begin{equation}
\int_{\Lambda }\frac{d^{4}k}{(2\pi )^{4}}\frac{1}{k^{2}-m^{2}}%
+p^{2}I_{log}(m^{2})+\,\,{\mbox{finite\,\,\, part}}\,\cdot  \label{26}
\end{equation}
The last term in equation (\ref{23}), namely the cross-terms, contain finite
contributions as well as ``nonlocal'' divergences. 
\begin{equation}
\Gamma _{nonlocal}=\Gamma _{fin}^{3}+\Gamma _{nonlocal}^{div}\cdot
\label{27}
\end{equation}
These nonlocal divergence contributions will always appear due to the
divergent subdiagrams (beyond two point functions) contained in the graph.
As we will show next in a particular example, the renormalization of
previous orders will always allow one to cancel these contributions if the
theory is renormalizable. In the present scheme the result is automatic and
follows from the operation we have just defined, in an algebraic manner.
There is no need for graphic representations of relevant contributions,
although it is possible.

The renormalized amplitude say, in $n^{th}$loop order, can therefore be
defined as 
\begin{eqnarray}
\Gamma _{R}^{(n)} &=&T^{D}\Gamma ^{(n)}-\Gamma _{local}^{div(n)}-\Gamma
_{nonlocal}^{div(n)} \\
&=&\Gamma _{fin}^{1(n)}+\Gamma _{fin}^{2(n)}+\Gamma _{fin}^{3(n)}  \nonumber
\end{eqnarray}
where the contributions $\Gamma _{local}^{div(n)}$ and $\Gamma
_{nonlocal}^{div(n)}$ contain the counterterms typical of\ order n as well
as the counterterms coming from divergent subdiagrams of previous order as
will become clear in the examples. Notice from the equation above that our
framework automatically delivers the counterterms 
\begin{equation}
\Gamma _{CT}^{1}=-\Gamma _{local}^{div}-\Gamma _{nonlocal}^{div}
\end{equation}
and just as in BPHZ, subtracting off the necessary counterterms leaves us
with the finite part of the amplitude. The main difference between our
method and BPHZ is that we never modify the original Feynman amplitude,
since we use an identity at the level of the integrand and BPHZ a Taylor
expansion. An immediate consequence of this difference is that in the
present procedure symmetries can always be preserved as has been shown in
references\cite{13}\cite{14}\cite{15}\cite{16}.

Now we proceed to evaluate the second class of diagrams, namely those which
contain two point functions as subdiagrams. Let us call U all the two point
diagrams contained in a given amplitude $\Gamma $. It is easy to see that
that they can be factored out inside of the total amplitude in the following
sense 
\begin{equation}
\Gamma =\prod_{all\ \ \Sigma _{j}\ \in \ U}{\cal R}_{j}\Sigma
_{j}^{(l)}(q_{j}^{2})
\end{equation}
where ${\cal R}_{j}$ stands for the remaining pieces in the amplitude, $j$%
characterizes a specific two point function, $q_{j}$ is one of the
integration momenta (but external to $\Sigma _{j}$). Now since the operation 
$T^{D}\Gamma $ is an identity, i.e. $T^{D}\Gamma =\Gamma $ we can define the
partially renormalized amplitude (with all two point function subdiagrams
properly renormalized ) as follows 
\begin{equation}
\bar{\Gamma}=\Gamma +\Gamma _{CT}^{2}
\end{equation}
therefore we have 
\begin{equation}
\Gamma _{CT}^{2}=\prod_{all\ \ \Sigma _{j}\ \in \ U}{\cal R}_{j}[\delta
_{j}^{(l)}m^{2}-A_{j}^{(l)}q_{j}^{2}]
\end{equation}
and $\Gamma _{CT}^{2}$ are all counterterms characteristic subdiagrams
involving two point functions. $\delta _{j}^{(l)}m^{2}$ stands for the mass
renormalization and $A_{j}^{(l)}$ for the wave function renormalization.
Explicit expressions for these objects will be given in the following
section where a specific example is worked out. In order to get the
renormalized amplitude of order n from $\bar{\Gamma}$ one proceeds in the
same way as for diagrams of class one defined above. We thus have 
\begin{eqnarray}
\Gamma _{R} &=&T^{D}\bar{\Gamma}-\bar{\Gamma}_{local}^{div}-\bar{\Gamma}%
_{nonlocal}^{div}  \nonumber \\
&=&\bar{\Gamma}_{fin}^{1}+\bar{\Gamma}_{fin}^{2}+\bar{\Gamma}_{fin}^{3}\cdot
\end{eqnarray}
The whole procedure will become apparent in the concrete example of the
following section

\section{$\protect\lambda \protect\phi _{6}^{3}$ Theory as an example}

Consider the $\lambda \phi _{6}^{3}$ theory Lagrangian,

\begin{equation}
{\cal L}=\frac{1}{2}\left[ \left( \partial _{\mu }\phi _{0}(x)\right)
^{2}-m_{0}^{2}\phi _{0}^{2}(x)\right] -\frac{\lambda _{0}}{3!}\phi
_{0}^{3}(x)  \label{a1}
\end{equation}
It is easy to show that a Feynman graph in this theory has the superficial
degree of divergence $D$ written as 
\begin{equation}
D=6-2N
\end{equation}
where $N$ is the number of external legs. This means that only Green's
functions with $N\leq 3$ are divergent. For the one-point functions we will
assume that we can impose the condition $\left\langle 0\right| \hat{\phi}%
\left| 0\right\rangle =0$ at all orders and we will not worry about
one-point diagrams. We will just work with the two and three-point Green's
functions which possess quadratic and logarithmic divergences.

We will effect the renormalization through the redefinition of the
Lagrangian parameters as: 
\begin{equation}
\phi _{0}=\sqrt{Z_{\phi }}\phi
\end{equation}
\begin{equation}
m_{0}^{2}=Z_{m}m^{2}
\end{equation}
\begin{equation}
\lambda _{0}=Z_{\lambda }\lambda
\end{equation}
which allow the Lagrangian to be rewritten as 
\begin{equation}
{\cal L}={{\cal L}_{F}}+{{\cal L}_{CT}}
\end{equation}
where 
\begin{equation}
{{\cal L}_{F}}=\frac{1}{2}\left[ \left( \partial _{\mu }\phi \right)
^{2}-m^{2}\phi ^{2}\right] -\frac{\lambda }{3!}\phi ^{3}
\end{equation}
and 
\begin{equation}
{\cal L}_{CT}=\frac{1}{2}\left[ (Z_{\phi }-1)\left( \partial _{\mu }\phi
\right) ^{2}-(Z_{\phi }Z_{m}-1)m^{2}\phi ^{2}\right] -(Z_{\phi
}^{3/2}Z_{\lambda }-1)\frac{\lambda }{3!}\phi ^{3}
\end{equation}
At the $n^{th}$ order one has 
\begin{equation}
{\cal L}_{CT}={\cal L}_{CT}^{(1)}+{\cal L}_{CT}^{(2)}+\ldots {\cal L}%
_{CT}^{(n)}
\end{equation}
We effect the renormalization at each order imposing the conditions:

\begin{itemize}
\item  Relative to the propagator 
\begin{equation}
D_{R}^{-1}(0)=-m^{2}
\end{equation}
and 
\begin{equation}
\left| \frac{\partial }{\partial p^{2}}D_{R}^{-1}(p^{2})\right| _{p=0}=1
\end{equation}

\item  Relative to vertex function 
\begin{equation}
-iM_{R}(0)=-i\lambda (1+{\rm finite\ corrections})
\end{equation}
We can rewrite the bare Lagrangian\ (\ref{a1}) as 
\begin{equation}
{\cal L}=\frac{1}{2}\left[ (1+A)\left( \partial _{\mu }\phi \right)
^{2}-(m^{2}+\delta m^{2})\phi ^{2}\right] -(1+B)\frac{\lambda }{3!}\phi
^{3}\,.
\end{equation}
\end{itemize}

in order to identify the renormalization constants 
\begin{equation}
Z_{\phi }^{(n)}=1+A^{(n)}
\end{equation}
\begin{equation}
Z_{m}^{(n)}=\frac{1}{Z_{\phi }^{(n)}m^{2}}(m^{2}+\delta ^{(n)}m^{2})
\end{equation}
\begin{equation}
Z_{\lambda }^{(n)}=\frac{1+B^{(n)}}{(Z_{\phi }^{(n)})^{3/2}}
\end{equation}
at each order by the imposing renormalization conditions. Since, in practice
we renormalize each diagram of the given order, the counterterms can be
written as 
\begin{equation}
B^{(n)}=\sum_{j=1}^{a}B_{j}^{(n)}
\end{equation}
\begin{equation}
A^{(n)}=\sum_{j=1}^{b}A_{j}^{(n)}
\end{equation}
\begin{equation}
\delta ^{(n)}m^{2}=\sum_{j=1}^{b}\delta _{j}^{(n)}m^{2}
\end{equation}
here $a,(b)$ is the number of three(two) point diagrams which contribute to
order $n$ .

At the $n^{th}$ order the inverse propagator function is written as 
\begin{eqnarray}
D_{R}^{-1}(p^{2}) &=&p^{2}-m^{2}-\Sigma _{R}^{(1)}(p^{2})-\Sigma
_{R}^{(2)}(p^{2})...  \nonumber \\
&&-\delta ^{(n)}m^{2}+A^{(n)}p^{2}-\Sigma ^{(n)}(p^{2})
\end{eqnarray}
and the vertex function as 
\begin{eqnarray}
-iM_{R}(p,p^{\prime }) &=&-i\lambda \{1+V_{R}^{(1)}(p,p^{\prime
})+V_{R}^{(2)}(p,p^{\prime })...  \nonumber \\
&&+V^{(n)}(p,p^{\prime })+B^{(n)}\}\cdot
\end{eqnarray}
Using the technique in each diagram contained in the $\Sigma ^{(n)}(p^{2})$
and in the $V^{(n)}(p,p^{\prime })$ amplitudes we separate the local
divergent part and identify all divergent substructures. Imposing
renormalization conditions we can always identify $A^{(n)}$, $\delta
^{(n)}m^{2}$ and $B^{(n)}$.

In order to identify the counterterms of the order in question and to write
the nonlocal ones in terms of divergences of lower orders, showing thus that
one needs not worry about them, it is convenient to define the following
functions:

\begin{itemize}
\item  Relative to vertex correction counterterms(type $j$ diagrams) 
\begin{eqnarray}
iB_{j}^{(n)} &=&(-i\lambda )^{2n+1}(i)^{3n}I_{\log 1}^{(n)}(m^{2},\Lambda ) 
\nonumber \\
&=&\Gamma _{local}^{div(n)}
\end{eqnarray}
where 
\begin{equation}
I_{\log 1}^{(n)}(m^{2},\Lambda )=\prod_{i=1}^{n}\int_{\Lambda }\frac{%
d^{6}k_{i}}{(2\pi )^{6}}\Upsilon ^{(n)}(k_{1},k_{2},...k_{n},m^{2})
\label{a2}
\end{equation}
with 
\begin{eqnarray}
&&\Upsilon ^{(n)}(k_{1},k_{2},...k_{n},m^{2})  \nonumber \\
&=&\frac{1}{(k_{1}^{2}-m^{2})^{3}}\left( \prod_{j=2}^{n}\frac{1}{%
(k_{j}^{2}-m^{2})^{2}}\right) Q(k_{i},k_{i+1},m^{2})\cdot
\end{eqnarray}
For $n=1$ 
\begin{equation}
Q=1,  \label{q1}
\end{equation}
otherwise 
\begin{equation}
Q=\prod_{i=1}^{n-1}\left\{ \frac{1}{[(k_{i}-k_{i+1})^{2}-m^{2}]}\right\}
\cdot  \label{q2}
\end{equation}
Notice that what we have defined here are generalizations of the simple
one-loop logarithmically divergent integral $I_{log}(m^{2})$ which we
encountered in our one-loop example.

\item  Relative to all finite contributions to vertex corrections(type $j$
diagram ), which corresponds to the renormalized diagram 
\begin{eqnarray}
\Gamma _{R}^{(n)} &=&\Gamma _{fin}^{1(n)}+\Gamma _{fin}^{3(n)}  \nonumber \\
&=&(-i\lambda )^{2n+1}(i)^{3n}\prod_{i=1}^{n}\int \frac{d^{6}k_{i}}{(2\pi
)^{6}}\Xi ^{(n)}(k_{1},...,k_{n},p,p^{\prime },m^{2})\cdot
\end{eqnarray}

\item  Relative to the finite contribution, defined in equation (\ref{24})
for the overlapping diagrams 
\begin{eqnarray}
\Gamma _{fin}^{1(n)} &=&(\Pi R)\prod_{i=1}^{n}f_{i}^{fin}(p,k_{i},m^{2}) 
\nonumber \\
&=&\frac{(-i\lambda )^{2n}(i)^{3n-1}}{2}\prod_{i=1}^{n}\int \frac{d^{6}k_{i}%
}{(2\pi )^{6}}\Theta ^{(n)}(k_{1},k_{2},...k_{n},p,m^{2})\cdot
\end{eqnarray}
\end{itemize}

In each order there will appear new types of divergent integrals. Therefore
throughout the text we will define some new divergent integrals similar to
the ones above (eq.(\ref{a2})). These quantities are always independent of
external momenta. Next we apply the procedure to all diagrams up to two
loops in order to exemplify how the method works. To $n^{th}$ order it
suffices to treat four cases, the first related to the vertex function and
the others to the self-energy, which contain the overlapping divergences,
two point functions as subdivergences and nested two point functions.

\subsection{Three point functions}

\subsubsection{The one loop order}

The vertex correction has only one contribution at one loop level whose
diagram is depicted in figure 1. The corresponding amplitude is 
\begin{eqnarray}
\Gamma &=&-iV^{(1)}(p,p^{\prime })  \nonumber \\
&=&\lambda ^{3}\int_{\Lambda }\frac{d^{6}k}{(2\pi )^{6}}\frac{1}{%
(k^{2}-m^{2})[(p-k)^{2}-m^{2}][(p^{\prime }-k)^{2}-m^{2}]}\cdot
\end{eqnarray}
Using the notation introduced in section 2 we write 
\begin{equation}
-iV^{(1)}(p,p^{\prime })=\int_{\Lambda }\frac{d^{6}k}{(2\pi )^{6}}%
R(k,m^{2},\lambda )f(p,k,m^{2})f(p^{\prime },k,m^{2})
\end{equation}
with 
\begin{equation}
R(k,m^{2},\lambda )=\frac{\lambda ^{3}}{(k^{2}-m^{2})}\cdot
\end{equation}
According to (IRT) we write, given that the divergence is logarithmic and
therefore $D=0$ 
\begin{equation}
-iT^{0}V^{(1)}(p,p^{\prime })=\Gamma _{local}^{div}+\Gamma _{fin}^{1}+\Gamma
_{fin}^{3}
\end{equation}
(recall that in this case $\Gamma _{fin}^{2}=\Gamma _{nonlocal}^{div}=0$ )
where 
\begin{eqnarray}
\Gamma _{local}^{div} &=&\int_{\Lambda }\frac{d^{6}k}{(2\pi )^{6}}%
R(k,m^{2},\lambda )f^{0}(k,m^{2})f^{0}(k,m^{2})  \nonumber \\
&=&\lambda ^{3}\int_{\Lambda }\frac{d^{6}k}{(2\pi )^{6}}\frac{1}{%
(k^{2}-m^{2})^{3}}  \nonumber \\
&=&\lambda ^{3}I_{\log 1}^{(1)}(m^{2},\Lambda )  \nonumber \\
&=&iB^{(1)}
\end{eqnarray}
and 
\begin{eqnarray}
\Gamma _{R} &=&\Gamma _{fin}^{1}+\Gamma _{fin}^{3}  \nonumber \\
&=&\lambda ^{3}\int \frac{d^{6}k}{(2\pi )^{6}}\Xi ^{(1)}(k,p,p^{\prime
},m^{2})
\end{eqnarray}
with 
\begin{eqnarray}
\lambda ^{3}\Xi ^{(1)}(k,p,p^{\prime },m^{2}) &=&R(k,m^{2},\lambda
)\{f^{fin}(k,p,m^{2})f^{fin}(k,p^{\prime },m^{2})  \nonumber \\
&+&f^{0}(k,m^{2})f^{fin}(k,p^{\prime },m^{2})  \nonumber \\
&&+f^{fin}(k,p,m^{2})f^{0}(k,m^{2})\}\cdot
\end{eqnarray}

Notice that the finite part of this diagram contains the cross-terms $%
f_{0}\cdot f_{fin}$ since its integral is finite.

\subsubsection{The two loop order}

Three diagram types contribute to the vertex correction at two loops. The
total amplitude can be written as 
\begin{equation}
-iV^{(2)}(p,p^{\prime })=-3iV_{1}^{(2)}(p,p^{\prime
})-iV_{2}^{(2)}(p,p^{\prime })-iV_{3}^{(2)}(p,p^{\prime })\cdot
\end{equation}
In this order the counterterms will be identified as 
\begin{equation}
B^{(n)}=3B_{1}^{(2)}+B_{2}^{(2)}+B_{3}^{(2)}\cdot
\end{equation}

The first amplitude $-iV_{1}^{(2)}(p,p^{\prime })$ corresponds to the
diagram in figure 2. This diagram contains a quadratic divergent subdiagram
(a first order two-point function correction). It can be completely
separated in terms of the internal momentum $k_{1}$ as mentioned before.
Then 
\begin{eqnarray}
\Gamma &=&-iV_{1}^{(2)}(p,p^{\prime })  \nonumber \\
&=&\int_{\Lambda }\frac{d^{6}k_{1}}{(2\pi )^{6}}\frac{(-i\lambda )^{3}(i)^{4}%
}{(k_{1}^{2}-m^{2})^{2}[(p-k_{1})^{2}-m^{2}][(p^{\prime }-k_{1})^{2}-m^{2}]}
\nonumber \\
&&\times \left\{ i\Sigma ^{(1)}(k_{1}^{2})\right\}  \label{c11}
\end{eqnarray}
where $i\Sigma ^{(1)}(k_{1}^{2})$ is the one loop self-energy amplitude. The
one loop renormalized self-energy is 
\begin{equation}
\Sigma _{R}^{(1)}(k_{1}^{2})=\Sigma _{CT}^{(1)}(k_{1}^{2})+\Sigma
^{(1)}(k_{1}^{2})
\end{equation}
where 
\begin{equation}
\Sigma _{CT}^{(1)}(k_{1}^{2})=\delta ^{(1)}m^{2}-A^{(1)}k_{1}^{2}\ \cdot
\end{equation}
Thus the amplitude containing no two point function substructure is directly
obtained as 
\begin{eqnarray}
\bar{\Gamma} &=&\Gamma +\Gamma _{CT}^{2}  \nonumber \\
&=&\int_{\Lambda }\frac{d^{6}k_{1}}{(2\pi )^{6}}\frac{(-i\lambda )^{3}(i)^{4}%
}{(k_{1}^{2}-m^{2})^{2}[(p-k_{1})^{2}-m^{2}][(p^{\prime }-k_{1})^{2}-m^{2}]}
\nonumber \\
&&\times \left\{ i\Sigma _{R}^{(1)}(k_{1}^{2})\right\} \cdot
\end{eqnarray}
All possible nonlocal divergences in this case will be canceled when we
consider the one loop renormalization. Next we use the IRT for the
logarithmic divergence. In our notation we obtain 
\begin{equation}
-iT^{0}\bar{\Gamma}=\bar{\Gamma}_{local}^{div}+\bar{\Gamma}_{fin}^{1}+\Gamma
_{fin}^{3}
\end{equation}
with 
\begin{equation}
\int \frac{d^{6}k_{2}}{(2\pi )^{6}}R(k_{1},k_{2},m^{2},\lambda )=\frac{%
(-i\lambda )^{3}(i)^{4}}{(k_{1}^{2}-m^{2})^{2}}\left\{ i\Sigma
_{R}^{(1)}(k_{1}^{2})\right\} =R(k_{1},m^{2},\lambda )\cdot
\end{equation}
The explicit expression for $i\Sigma _{R}^{(1)}(k_{1}^{2})$ will be given in
equation (\ref{b1}). Using the above expression for $R(k_{1},m^{2},\lambda )$
\begin{eqnarray}
\bar{\Gamma}_{local}^{div} &=&\int_{\Lambda }\frac{d^{6}k_{1}}{(2\pi )^{6}}%
R(k_{1},m^{2},\lambda )f^{0}(k_{1},m^{2})f^{0}(k_{1},m^{2})  \nonumber \\
&=&(-i\lambda )^{3}(i)^{4}I_{\log 2}^{(2)}(m^{2},\lambda ^{2},\Lambda
)=iB_{1}^{(2)}\cdot
\end{eqnarray}
We have just defined another logarithmic divergent quantity which is
characteristic of the two-loop order. Note the explicit appearance of
coupling constant. This should emphasize the fact that the amplitude depends
on a two point function subdiagram, which has been properly renormalized.
All counterterms possessing such type of subdiagram will look like this. 
\begin{equation}
I_{\log 2}^{(2)}(m^{2},\lambda ^{2},\Lambda )=\int_{\Lambda }\frac{d^{6}k_{1}%
}{(2\pi )^{6}}\frac{1}{(k_{1}^{2}-m^{2})^{4}}\left\{ i\Sigma
_{R}^{(1)}(k_{1}^{2})\right\} .  \label{a4}
\end{equation}
The finite part is 
\begin{eqnarray}
\bar{\Gamma}_{fin}^{1}+\bar{\Gamma}_{fin}^{3} &=&\int \frac{d^{6}k_{1}}{%
(2\pi )^{6}}R(k_{1},m^{2},\lambda )\{  \nonumber \\
&&f^{0}(k_{1},m^{2})f^{fin}(p^{\prime },k_{1},m^{2})  \nonumber \\
&&+f^{fin}(p,k_{1},m^{2})f^{0}(k_{1},m^{2})  \nonumber \\
&&+f^{fin}(p,k_{1},m^{2})f^{fin}(p^{\prime },k_{1},m^{2})\}\cdot
\end{eqnarray}
It is not necessary to give explicit expressions for the finite part and
therefore we make explicit the divergent contributions only.

Now we consider the diagram corresponding to the second contribution $%
-iV_{2}^{(2)}(p,p^{\prime })$ which belong to class one (figure 3). The
amplitude reads 
\begin{eqnarray}
-iV_{2}^{(2)}(p,p^{\prime }) &=&\int_{\Lambda }\frac{d^{6}k_{1}}{(2\pi )^{6}}%
\int_{\Lambda }\frac{d^{6}k_{2}}{(2\pi )^{6}}R(k_{1},k_{2},m^{2},\lambda ) 
\nonumber \\
&&\times f(p,k_{1},m^{2})f(p^{\prime },k_{1},m^{2})f(p^{\prime },k_{2},m^{2})
\end{eqnarray}
with 
\begin{equation}
R(k_{1},k_{2},m^{2},\lambda )=\frac{(-i\lambda )^{5}(i)^{6}}{%
(k_{1}^{2}-m^{2})(k_{2}^{2}-m^{2})[(k_{1}-k_{2})^{2}-m^{2}]}\cdot
\end{equation}
Using the IRT we have 
\begin{equation}
-iT^{0}\,\,V_{1}^{(2)}(p,p^{\prime })=\Gamma _{local}^{div}+\Gamma
_{fin}^{1}+\Gamma _{nonlocal}
\end{equation}
where 
\begin{eqnarray}
\Gamma _{local}^{div} &=&(-i\lambda )^{5}(i)^{6}I_{\log
1}^{(2)}(m^{2},\Lambda )=iB_{2}^{(2)}  \nonumber \\
&=&\int_{\Lambda }\frac{d^{6}k_{1}}{(2\pi )^{6}}\int_{\Lambda }\frac{%
d^{6}k_{2}}{(2\pi )^{6}}\frac{(-i\lambda )^{5}(i)^{6}}{%
(k_{1}^{2}-m^{2})^{3}(k_{2}^{2}-m^{2})^{2}[(k_{1}-k_{2})^{2}-m^{2}]} 
\nonumber \\
&=&(-i\lambda )^{5}(i)^{6}\int_{\Lambda }\frac{d^{6}k_{1}}{(2\pi )^{6}}%
\int_{\Lambda }\frac{d^{6}k_{2}}{(2\pi )^{6}}\Upsilon
^{(2)}(k_{1},k_{2},m^{2})\cdot
\end{eqnarray}
In this type of structure (to all orders) the nonlocal contribution $\Gamma
_{nonlocal}$ will have the form 
\begin{equation}
\Gamma _{nonlocal}=\Gamma _{fin}^{3}+\Gamma _{nonlocal}^{div}
\end{equation}
and in this case we have 
\begin{equation}
\Gamma _{nonlocal}^{div}=(-i\lambda )^{5}(i)^{6}\int =\frac{d^{6}k_{1}}{%
(2\pi )^{6}}\int_{\Lambda }\frac{d^{6}k_{2}}{(2\pi )^{6}}\Xi
^{(1)}(k_{1},p,p^{\prime },m^{2})\Upsilon ^{(1)}(k_{2},m^{2})\cdot
\end{equation}
Note that this term is completely written in terms of one loop contributions
already considered. Therefore it poses no problem to renormalization. This
particular example illustrates a basic difference between the present method
and others: the subdivergences need not be previously identified. They
appear algebraically. In cases were it is simple to identify the
subdivergences , this is not necessarily a great advantage. However in
higher orders it might become considerably simpler to identify all divergent
substructures in an algebraic fashion. In fact, as will become clear in what
follows, the procedure is designed to display all relevant (to
renormalization) subdivergences. The finite contributions can be written as 
\begin{eqnarray}
\Gamma _{R}^{(2)} &=&\Gamma _{fin}^{1}+\Gamma _{fin}^{3}  \nonumber \\
&=&(-i\lambda )^{5}(i)^{6}\int \frac{d^{6}k_{1}}{(2\pi )^{6}}\int \frac{%
d^{6}k_{2}}{(2\pi )^{6}}\Xi ^{(2)}(k_{1},k_{2},p,p^{\prime },m^{2})
\end{eqnarray}
with 
\[
\Xi =(-i\lambda )^{5}(i)^{6}\Xi ^{(2)}(k_{1},k_{2},p,p^{\prime },m^{2}) 
\]
\begin{eqnarray}
\Xi &=&R(k_{1},k_{2},m^{2},\lambda )  \nonumber \\
&&\times \{f^{fin}(k_{1},p,m^{2})f^{fin}(k_{1},p^{\prime
},m^{2})f^{fin}(k_{2},p^{\prime },m^{2})  \nonumber \\
&&\left. +f^{0}(k_{1},m^{2})f^{0}(k_{1},m^{2})f^{fin}(k_{2},p^{\prime
},m^{2})\right.  \nonumber \\
&&+f^{0}(k_{1},m^{2})f^{fin}(k_{1},p^{\prime },m^{2})f^{fin}(k_{2},p^{\prime
},m^{2})  \nonumber \\
&&+f^{fin}(k_{1},p,m^{2})f^{0}(k_{1},m^{2})f^{fin}(k_{2},p^{\prime },m^{2})\}
\nonumber \\
&&+\frac{(-i\lambda )^{5}(i)^{6}(2k_{1}.k_{2}-k_{1}^{2})\Xi
^{(1)}(k_{1},p,p^{\prime },m^{2})}{%
(k_{2}^{2}-m^{2})^{3}[(k_{1}-k_{2})^{2}-m^{2}]}\cdot
\end{eqnarray}
The last term in the above equation is obtained by using the operation (\ref
{12}) considering $k_{1}$ as external momentum. This is necessary to
identify the one loop structure.

The last two loop diagram $-iV_{3}^{(2)}(p,p^{\prime })$ is depicted in the
figure 4. The corresponding amplitude is 
\begin{eqnarray}
-iV_{3}^{(2)}(p,p^{\prime }) &=&\int_{\Lambda }\frac{d^{6}k_{1}}{(2\pi )^{6}}%
\int_{\Lambda }\frac{d^{6}k_{2}}{(2\pi )^{6}}\{R(k_{1},k_{2},m^{2},\lambda
)f(p,k_{1},m^{2})  \nonumber \\
&&\times f(p^{\prime },k_{2},m^{2})f(p-p^{\prime },k_{1}-k_{2},m^{2})\}
\label{68}
\end{eqnarray}
with 
\begin{equation}
R(k_{1},k_{2},m^{2},\lambda )=\frac{(-i\lambda )^{5}(i)^{6}}{%
(k_{1}^{2}-m^{2})(k_{2}^{2}-m^{2})[(k_{1}-k_{2})^{2}-m^{2}]}\cdot  \label{69}
\end{equation}
Using the IRT we have 
\begin{equation}
-iT^{0}V_{1}^{(2)}(p,p^{\prime })=\Gamma _{local}^{div}+\Gamma _{R}
\label{70}
\end{equation}
where 
\begin{eqnarray}
\Gamma _{local}^{div} &=&i\lambda ^{5}\int_{\Lambda }\frac{d^{6}k_{1}}{(2\pi
)^{6}}\int_{\Lambda }\frac{d^{6}k_{2}}{(2\pi )^{6}}\frac{1}{%
(k_{1}^{2}-m^{2})^{2}(k_{2}^{2}-m^{2})^{2}[(k_{1}-k_{2})^{2}-m^{2}]^{2}} 
\nonumber \\
&=&i\lambda ^{5}I_{\log 3}^{(2)}(m^{2},\Lambda )=iB_{3}^{(2)}\cdot
\label{71}
\end{eqnarray}
We defined above another logarithmic divergent quantity. This diagram type
is often called a primitively divergent diagram. Note that there are no
subdivergences.

\subsubsection{The $n$-loop order}

As discussed before we now consider only one contribution of each kind. The
vertex type contribution depicted in figure 5 is the first one. It will
appear as a substructure of the overlapping self-energy diagram which we
will also consider.

The amplitude corresponding to the vertex correction in figure 5 is 
\begin{equation}
-iV_{1}^{(n)}(p,p^{\prime })=(\Pi R)(\Pi f)  \label{72}
\end{equation}
with 
\begin{equation}
(\Pi R)=(-i\lambda )^{2n+1}(i)^{3n}\left\{ \prod_{j=1}^{n}\int_{\Lambda }%
\frac{d^{6}k_{j}}{(2\pi )^{6}}\frac{1}{k_{j}^{2}-m^{2}}\right\}
Q(k_{i},k_{i+1},m^{2})  \label{73}
\end{equation}
where $Q$ is the same function as defined in (\ref{q1}) and (\ref{q2}). The
subscript $1$ in $V_{1}^{(n)}(p,p^{\prime })$ refers to the fact that only
one diagram is being considered (type $1$ ). The external momentum dependent
part $(\Pi f)$ is given by 
\begin{equation}
(\Pi f)=\left\{ \frac{1}{(p-k_{1})^{2}-m^{2}}\right\} \prod_{j=1}^{n}\left\{ 
\frac{1}{(p^{\prime }-k_{j})^{2}-m^{2}}\right\} \cdot  \label{74}
\end{equation}
Using the IRT, we get 
\begin{eqnarray}
T^{0}(\Pi f) &=&\{f^{0}(k_{1},m^{2})+f^{fin}(k_{1},p,m^{2})\}  \nonumber \\
&&\times \prod_{j=1}^{n}\{f^{0}(k_{j},m^{2})+f^{fin}(k_{j},p^{\prime
},m^{2})\}\cdot  \label{75}
\end{eqnarray}
In the same way we have 
\begin{equation}
-iT^{0}V_{1}^{(n)}(p,p^{\prime })=\Gamma _{local}+\Gamma _{fin}^{1}+\Gamma
_{nonlocal}\cdot  \label{76}
\end{equation}
Since $\Gamma _{fin}^{2}=0$ we can write 
\begin{eqnarray}
\Gamma _{local}^{div} &=&iB_{1}^{(n)}  \nonumber \\
&=&(\Pi R)f_{1}^{0}(k_{1},m^{2})\prod_{j=1}^{n}f_{j}^{0}(k_{j},m^{2}) 
\nonumber \\
&=&(-i\lambda )^{2n+1}(i)^{3n}I_{\log 1}^{(n)}(m^{2},\Lambda )  \nonumber \\
&=&(-i\lambda )^{2n+1}(i)^{3n}\prod_{j=1}^{n}\int_{\Lambda }\frac{d^{6}k_{j}%
}{(2\pi )^{6}}\Upsilon ^{(n)}(k_{1},k_{2},...k_{n},m^{2})
\end{eqnarray}
and 
\begin{equation}
\Gamma _{nonlocal}=\Gamma _{fin}^{3}+\Gamma _{nonlocal}^{div}  \label{78}
\end{equation}
where 
\begin{eqnarray}
\Gamma _{nonlocal}^{div} &=&(-i\lambda
)^{2n+1}(i)^{3n}\prod_{j=1}^{n}\int_{\Lambda }\frac{d^{6}k_{j}}{(2\pi )^{6}}
\nonumber \\
&&\sum_{a=1}^{n-1}\Xi ^{(a)}(k_{1},..,k_{a};p,p^{\prime },m^{2})\Upsilon
^{(n-a)}(k_{a+1},..,k_{n};m^{2})\cdot
\end{eqnarray}
Note that the above equation contains subdivergences which have already
appeared at lower orders and have already been included in the Lagrangian.
Here we clearly see that the application of the method displays all the
subdivergencies in an algebraic way. Moreover it stresses the inductive
character of the method. If we assume that the theory is renormalized at $%
(n-1)^{th}$ order, the contribution at $n^{th}$ order will solely depend on
structures (finite and divergent) which have already played their role at
lower orders. Also it is noteworthy that {\it all} divergencies and finite
parts of {\it all} previous orders play an important role at $n^{th}$ order.
The above expression clearly displays a difference between this method and
BPHZ. In BPHZ one would have to deal with the detailed topology of the
diagram first. Here the counterterms appear algebraically. This simplicity
may be due to the example we are working with. However, it constitutes even
in this case a notorious simplification.

\subsection{Two-point functions}

\subsubsection{The one loop order}

The self-energy has only one diagram contribution at one loop level which we
identify in figure 6. It corresponds to the amplitude 
\begin{equation}
i\Sigma ^{(1)}(p^{2})=\frac{\lambda ^{2}}{2}\int_{\Lambda }\frac{d^{6}k}{%
(2\pi )^{6}}\frac{1}{(k^{2}-m^{2})[(p-k)^{2}-m^{2}]}  \label{80}
\end{equation}
where 
\begin{equation}
R(k,m^{2},\lambda )=\frac{\lambda ^{2}}{2}\frac{1}{(k^{2}-m^{2})}\cdot
\label{81}
\end{equation}
Using IRT we have 
\begin{equation}
(T^{2})i\Sigma ^{(1)}(p^{2})=\Gamma _{local}^{div}+\Gamma _{fin}^{1}
\label{82}
\end{equation}
with 
\begin{equation}
\Gamma _{local}^{div}=\frac{\lambda ^{2}}{2}\{I_{quad}^{(1)}(m^{2},\Lambda
)+p^{2}[g_{\mu \nu }\frac{4}{6}I_{\log }^{\mu \nu (1)}(m^{2},\Lambda
)-I_{\log 1}^{(1)}(m^{2},\Lambda )]\},  \label{83}
\end{equation}
where we have defined 
\begin{equation}
I_{quad}^{(1)}(m^{2},\Lambda )=\int_{\Lambda }\frac{d^{6}k}{(2\pi )^{6}}%
\frac{1}{(k^{2}-m^{2})^{2}}  \label{84}
\end{equation}
and 
\begin{equation}
I_{\log }^{\mu \nu (1)}(m^{2},\Lambda )=\int_{\Lambda }\frac{d^{6}k}{(2\pi
)^{6}}\frac{k^{\mu }k^{\nu }}{(k^{2}-m^{2})^{4}}\cdot  \label{85}
\end{equation}
The finite part is 
\begin{eqnarray}
\Gamma _{fin}^{1} &=&\Gamma _{R}=\frac{\lambda ^{2}}{2}\int \frac{d^{6}k}{%
(2\pi )^{6}}\Theta ^{(1)}(k,p,m^{2})  \nonumber \\
&=&\frac{\lambda ^{2}}{2}\left\{ \int \frac{d^{6}k}{(2\pi )^{6}}\frac{(p^{4})%
}{(k^{2}-m^{2})^{4}}\right.  \nonumber \\
&&\left. +\int \frac{d^{6}k}{(2\pi )^{6}}\frac{(2p.k-p^{2})^{3}}{%
(k^{2}-m^{2})^{4}[(p-k)^{2}-m^{2}]}\right\} \cdot  \label{86}
\end{eqnarray}
The explicit calculation of the integral in the above equation leads to 
\begin{equation}
\Gamma _{R}=i\Sigma _{R}^{(1)}(p^{2})=\frac{\lambda ^{2}}{4}\frac{i}{(4\pi
)^{3}}\left\{ (p^{2}-3m^{2})F(m^{2},p^{2})-\frac{p^{2}}{2}\right\}
\label{b1}
\end{equation}
where $F(m^{2},p^{2}),$\ for $p^{2}<4m^{2}$ ,is given by 
\begin{equation}
F(m^{2},p^{2})=\frac{\sqrt{4m^{2}-p^{2}}}{\left| p\right| }\left[ 2\arctan
\left( \frac{\sqrt{4m^{2}-p^{2}}}{\left| p\right| }\right) +\pi \right] -2
\label{88}
\end{equation}
and for $p^{2}>4m^{2}$ , 
\begin{equation}
F(m^{2},p^{2})=-\frac{\sqrt{p^{2}-4m^{2}}}{\left| p\right| }\left[ \ln
\left( \frac{\left| p\right| -\sqrt{p^{2}-4m^{2}}}{\left| p\right| +\sqrt{%
p^{2}-4m^{2}}}\right) +i\pi \right] -2\ \cdot  \label{89}
\end{equation}
We now summarize the results obtained so far for one loop the
renormalization, 
\begin{equation}
A^{(1)}=\frac{i\lambda ^{2}}{2}\left\{ I_{\log 1}^{(1)}(m^{2},\Lambda )-%
\frac{4}{6}g_{\mu \nu }I_{\log }^{\mu \nu (1)}(m^{2},\Lambda )\right\}
\label{125}
\end{equation}
\begin{equation}
\delta ^{(1)}m^{2}=\frac{\lambda ^{2}}{2}iI_{quad}^{(1)}(m^{2},\Lambda )
\label{126}
\end{equation}
and 
\begin{equation}
B^{(1)}=-i\lambda ^{2}I_{\log 1}^{(1)}(m^{2},\Lambda )  \label{127}
\end{equation}
where $I_{\log 1}^{(1)}(m^{2},\Lambda )$ , $I_{\log }^{\mu \nu
(1)}(m^{2},\Lambda )$ and $I_{quad}^{(1)}(m^{2},\Lambda )$ \ are defined in (%
\ref{a2}), (\ref{85}) and (\ref{84}), respectively.

\subsubsection{The two loop order}

Two types of diagram contribute to the self energy correction at two loops.
The total amplitude can be written as 
\begin{equation}
\Sigma ^{(2)}(p^{2})=2\Sigma _{1}^{(2)}(p^{2})+\Sigma _{2}^{(2)}(p^{2})\cdot
\label{90}
\end{equation}
Here the counterterms to be identified are 
\begin{equation}
A^{(2)}=2A_{1}^{(2)}+A_{2}^{(2)}
\end{equation}
\begin{equation}
\delta ^{(2)}m^{2}=2\delta _{1}^{(2)}m^{2}+\delta _{2}^{(2)}m^{2}\cdot
\end{equation}
The first amplitude $i\Sigma _{1}^{(2)}(p^{2})$ corresponds to the diagram
in figure 7. This is the same case we have seen in equation (\ref{c11}).
Considering the one loop renormalization we can write 
\begin{equation}
\bar{\Gamma}=i\bar{\Sigma}_{1}^{(2)}(p^{2})=\int_{\Lambda }\frac{d^{6}k_{1}}{%
(2\pi )^{6}}\frac{(-i\lambda )^{2}(i)^{3}}{%
(k_{1}^{2}-m^{2})^{2}[(p-k_{1})^{2}-m^{2}]}\{i\Sigma
_{R}^{(1)}(k_{1}^{2})\}\cdot  \label{91}
\end{equation}
Then we apply IRT and obtain 
\begin{equation}
iT^{2}\bar{\Sigma}_{1}^{(2)}(p^{2})=\bar{\Gamma}_{local}^{div}+\bar{\Gamma}%
_{fin}^{1}  \label{92}
\end{equation}
in which 
\begin{eqnarray}
\bar{\Gamma}_{local}^{div} &=&\lambda ^{2}i\left\{
I_{quad2}^{(2)}(m^{2},\lambda ^{2},\Lambda )\right.  \nonumber \\
&&\left. +p^{2}[\frac{4g_{\mu \nu }}{6}I_{\log 2}^{\mu \nu
(2)}(m^{2},\lambda ^{2},\Lambda )-I_{\log 2}^{(2)}(m^{2},\lambda
^{2},\Lambda )]\right\}  \label{93}
\end{eqnarray}
and 
\begin{equation}
I_{quad2}^{(2)}(m^{2},\lambda ^{2},\Lambda )=\int_{\Lambda }\frac{d^{6}k_{1}%
}{(2\pi )^{6}}\frac{1}{(k_{1}^{2}-m^{2})^{3}}\left\{ i\Sigma
_{R}^{(1)}(k_{1}^{2})\right\}  \label{94}
\end{equation}
\begin{equation}
I_{\log 2}^{\mu \nu (2)}(m^{2},\lambda ^{2},\Lambda )=\int_{\Lambda }\frac{%
d^{6}k_{1}}{(2\pi )^{6}}\frac{k^{\mu }k^{\nu }}{(k_{1}^{2}-m^{2})^{5}}%
\left\{ i\Sigma _{R}^{(1)}(k_{1}^{2})\right\}  \label{95}
\end{equation}
whereas 
\begin{eqnarray}
\bar{\Gamma}_{fin}^{1} &=&\lambda ^{2}i\int \frac{d^{6}k_{1}}{(2\pi )^{6}}%
\left\{ \frac{p^{4}}{(k_{1}^{2}-m^{2})^{5}}\right.  \nonumber \\
&&\left. +\frac{(2p.k_{1}-p^{2})^{3}}{%
(k_{1}^{2}-m^{2})^{5}[(p-k_{1})^{2}-m^{2}]}\right\} \left\{ i\Sigma
_{R}^{(1)}(k_{1}^{2})\right\} \cdot  \label{96}
\end{eqnarray}

The second amplitude $i\Sigma _{2}^{(2)}(p^{2})$ corresponds to the diagram
in figure 8. It reads 
\begin{equation}
i\Sigma _{2}^{(2)}(p^{2})=\int_{\Lambda }\frac{d^{6}k_{1}}{(2\pi )^{6}}%
\int_{\Lambda }\frac{d^{6}k_{2}}{(2\pi )^{6}}R(k_{1},k_{2},m^{2},\lambda
)f(p,k_{1},m^{2})f(p,k_{2},m^{2})  \label{97}
\end{equation}
with 
\begin{equation}
R(k_{1},k_{2},m^{2},\lambda )=\frac{1}{2}\{\frac{(-i\lambda )^{4}(i)^{5}}{%
(k_{1}^{2}-m^{2})(k_{2}^{2}-m^{2})[(k_{1}-k_{2})^{2}-m^{2}]}\}\cdot
\label{98}
\end{equation}
The same procedure enables us to write

\begin{eqnarray}
\Gamma _{local}^{div} &=&\frac{i\lambda ^{4}}{2}\left\{
I_{quad1}^{(2)}(m^{2},\Lambda )\right.  \nonumber \\
&&\left. +2p^{2}[\frac{4g_{\mu \nu }}{6}I_{\log 1}^{\mu \nu
(2)}(m^{2},\Lambda )-I_{\log 1}^{(2)}(m^{2},\Lambda )]\right\}
\end{eqnarray}
where 
\begin{eqnarray}
I_{quad1}^{(2)}(m^{2},\Lambda ) &=&\int_{\Lambda }\frac{d^{6}k_{1}}{(2\pi
)^{6}}\int_{\Lambda }\frac{d^{6}k_{2}}{(2\pi )^{6}}  \nonumber \\
&&\times \frac{1}{%
(k_{1}^{2}-m^{2})^{2}(k_{2}^{2}-m^{2})^{2}[(k_{1}-k_{2})^{2}-m^{2}]}
\label{100}
\end{eqnarray}
\begin{eqnarray}
I_{\log 1}^{\mu \nu (2)}(m^{2},\Lambda ) &=&\int_{\Lambda }\frac{d^{6}k_{1}}{%
(2\pi )^{6}}\int_{\Lambda }\frac{d^{6}k_{2}}{(2\pi )^{6}}  \nonumber \\
&&\times \frac{k_{2}^{\mu }k_{2}^{\nu }}{%
(k_{1}^{2}-m^{2})^{2}(k_{2}^{2}-m^{2})^{4}[(k_{1}-k_{2})^{2}-m^{2}]}
\label{a3}
\end{eqnarray}
and the finite part coming from this contribution is 
\begin{equation}
\Gamma _{fin}^{2}=\frac{i\lambda ^{4}}{2}(I_{1}+2I_{2}+I_{3})  \label{102}
\end{equation}
with 
\begin{equation}
I_{1}=\int \frac{d^{6}k_{1}}{(2\pi )^{6}}\int \frac{d^{6}k_{2}}{(2\pi )^{6}}%
\frac{p^{4}}{%
(k_{1}^{2}-m^{2})^{3}(k_{2}^{2}-m^{2})^{3}[(k_{1}-k_{2})^{2}-m^{2}]}
\label{103}
\end{equation}
\begin{equation}
I_{2}=\int \frac{d^{6}k_{1}}{(2\pi )^{6}}\int \frac{d^{6}k_{2}}{(2\pi )^{6}}%
\frac{-4p^{2}(p.k_{2})^{2}}{%
(k_{1}^{2}-m^{2})^{3}(k_{2}^{2}-m^{2})^{4}[(k_{1}-k_{2})^{2}-m^{2}]}
\end{equation}
\begin{equation}
I_{3}=\int \frac{d^{6}k_{1}}{(2\pi )^{6}}\int \frac{d^{6}k_{2}}{(2\pi )^{6}}%
\frac{16(p.k_{1})^{2}(p.k_{2})^{2}}{%
(k_{1}^{2}-m^{2})^{4}(k_{2}^{2}-m^{2})^{4}[(k_{1}-k_{2})^{2}-m^{2}]}
\label{105}
\end{equation}

\begin{equation}
\Gamma _{nonlocal}=\Gamma _{fin}^{3}+\Gamma _{nonlocal}^{div}\cdot
\label{106}
\end{equation}
In terms of functions $\Theta ^{(1)}(k_{i},p,m^{2})$ and $\Upsilon
^{(1)}(k_{i},m^{2})$ we can write 
\begin{eqnarray}
\Gamma _{nonlocal}^{div} &=&\frac{(-i\lambda )^{4}(i)^{5}}{2}\int_{\Lambda }%
\frac{d^{6}k_{1}}{(2\pi )^{6}}\int_{\Lambda }\frac{d^{6}k_{2}}{(2\pi )^{6}}\{
\nonumber \\
&&\Theta ^{(1)}(k_{1},p,m^{2})\Upsilon ^{(1)}(k_{2},m^{2})  \nonumber \\
&&+\Theta ^{(1)}(k_{2},p,m^{2})\Upsilon ^{(1)}(k_{1},m^{2})\}\cdot
\label{107}
\end{eqnarray}
Note that $\Upsilon ^{(1)}$ is (the integrand of a) logarithmic divergence,
which, in DR would give us $1/\epsilon $ and when multiplied by the
remaining pieces of the amplitude would produce the celebrated term $\ln
p^{2}/\epsilon $ \cite{17}. The (other) finite contributions are 
\begin{equation}
\Gamma _{fin}^{1}=\frac{(-i\lambda )^{4}(i)^{5}}{2}\int \frac{d^{6}k_{1}}{%
(2\pi )^{6}}\int \frac{d^{6}k_{2}}{(2\pi )^{6}}\Theta
^{(2)}(k_{1},k_{2},p,m^{2})  \label{108}
\end{equation}
and 
\begin{eqnarray}
\Gamma _{fin}^{3} &=&\frac{(-i\lambda )^{4}(i)^{5}}{2}\int \frac{d^{6}k_{1}}{%
(2\pi )^{6}}\int \frac{d^{6}k_{2}}{(2\pi )^{6}}\{  \nonumber \\
&&\Theta ^{(1)}(k_{1},p,m^{2})\frac{2k_{1}.k_{2}-k_{1}^{2}}{%
(k_{2}^{2}-m^{2})^{3}[(k_{1}-k_{2})^{2}-m^{2}]}  \nonumber \\
&&+\Theta ^{(1)}(k_{2},p,m^{2})\frac{2k_{1}.k_{2}-k_{2}^{2}}{%
(k_{1}^{2}-m^{2})^{3}[(k_{1}-k_{2})^{2}-m^{2}]}  \nonumber \\
&&+\int \frac{d^{6}k_{1}}{(2\pi )^{6}}\int \frac{d^{6}k_{2}}{(2\pi )^{6}}%
R(k_{1},k_{2},m^{2},\lambda )\{  \nonumber \\
&&f^{fin}(p,k_{1},m^{2})[f_{1}(k_{2},m^{2})+f_{2}(k_{2},m^{2})]  \nonumber \\
&&+f^{fin}(p,k_{2},m^{2})[f_{1}(k_{1},m^{2})+f_{2}(k_{1},m^{2})]\}\}\cdot
\label{109}
\end{eqnarray}
Summarizing the two loop renormalization constants obtained are, 
\begin{eqnarray}
A^{(2)} &=&\frac{\lambda ^{4}}{2}[\frac{4}{6}g_{\mu \nu }I_{\log 1}^{\mu \nu
(2)}(m^{2},\Lambda )-I_{\log 1}^{(2)}(m^{2},\Lambda )]  \nonumber \\
&&+\lambda ^{2}[\frac{4}{6}g_{\mu \nu }I_{\log 2}^{\mu \nu
(2)}(m^{2},\lambda ^{2},\Lambda )-I_{\log 2}^{(2)}(m^{2},\lambda
^{2},\Lambda )]  \label{128}
\end{eqnarray}
\begin{equation}
\delta ^{(2)}m^{2}=-[\frac{\lambda ^{4}}{2}I_{quad1}^{(2)}(m^{2},\Lambda
)+\lambda ^{2}I_{quad2}^{(2)}(m^{2},\lambda ^{2},\Lambda )]  \label{129}
\end{equation}
and 
\begin{equation}
B^{(2)}=\lambda ^{4}[I_{\log 1}^{(2)}(m^{2},\Lambda )+I_{\log
3}^{(2)}(m^{2},\Lambda )]+3\lambda ^{2}I_{\log 2}^{(2)}(m^{2},\lambda
^{2},\Lambda )  \label{130}
\end{equation}
where 
\[
I_{\log 1}^{(2)}(m^{2},\Lambda ),I_{\log 1}^{\mu \nu (2)}(m^{2},\Lambda
),I_{\log 2}^{(2)}(m^{2},\lambda ^{2},\Lambda ),I_{\log 2}^{\mu \nu
(2)}(m^{2},\lambda ^{2},\Lambda ),I_{\log 3}^{(2)}(m^{2},\Lambda ) 
\]
and 
\[
I_{quad1}^{(2)}(m^{2},\Lambda ),I_{quad2}^{(2)}(m^{2}.\lambda ^{2},\Lambda ) 
\]
are defined in equations (\ref{a2}), (\ref{a3}), (\ref{a4}), (\ref{95}), (%
\ref{71}), (\ref{100}) and (\ref{94}) respectively.

\subsubsection{The $n$-loop order}

Let us first consider the overlapping self energy diagram of figure 9. It
corresponds to the amplitude 
\begin{equation}
i\Sigma _{1}^{(n)}(p^{2})=(\Pi R)(\Pi f)  \label{110}
\end{equation}
with 
\begin{equation}
(\Pi R)=\frac{(-i\lambda )^{2n}(i)^{3n-1}}{2}\left\{
\prod_{j=1}^{n}\int_{\Lambda }\frac{d^{6}k_{j}}{(2\pi )^{6}}\frac{1}{%
k_{j}^{2}-m^{2}}\right\} Q(k_{i},k_{i+1},m^{2})\cdot  \label{111}
\end{equation}
The external momentum dependent part is 
\begin{equation}
(\Pi f)=\prod_{j=1}^{n}\frac{1}{[(p-k_{j})^{2}-m^{2}]}\cdot  \label{112}
\end{equation}
Using the technique we have, as usual 
\begin{equation}
iT^{2}\Sigma _{1}^{(n)}(p^{2})=\Gamma _{fin}^{1}+\Gamma _{local}+\Gamma
_{nonlocal}  \label{113}
\end{equation}
where 
\begin{equation}
\Gamma _{fin}^{1}=\frac{(-i\lambda )^{2n}(i)^{3n-1}}{2}\prod_{j=1}^{n}\int 
\frac{d^{6}k_{j}}{(2\pi )^{6}}\Theta ^{(n)}(k_{1},k_{2},...k_{n};p,m^{2})
\label{114}
\end{equation}
and 
\begin{eqnarray}
\Gamma _{local} &=&\Gamma _{fin}^{2}+\Gamma _{local}^{div}  \nonumber \\
&=&(\Pi
R)\prod_{j=1}^{n}%
\{f^{0}(k_{j},m^{2})+f^{1}(k_{j},m^{2},p)+f^{2}(k_{j},m^{2},p)\}\cdot
\end{eqnarray}
In this case we have $\Gamma _{fin}^{2}\neq 0$ . The counterterms
characteristic of the $n^{th}$ order are identified in the equation 
\[
\Gamma _{local}^{div}=i(\delta _{1}^{(n)}m^{2}-A_{1}^{(n)}p^{2})\cdot 
\]
The nonlocal part is 
\begin{equation}
\Gamma _{nonlocal}=\Gamma _{fin}^{3}+\Gamma _{nonlocal}^{div}  \label{116}
\end{equation}
where 
\begin{eqnarray}
\Gamma _{nonlocal}^{div} &=&\frac{(-i\lambda )^{2n}(i)^{3n-1}}{2}%
\prod_{j=1}^{n}\int_{\Lambda }\frac{d^{6}k_{j}}{(2\pi )^{6}}  \label{145} \\
&&\Bigg\{\sum_{a=1}^{n-1}\Theta
^{(a)}(k_{1},k_{2},..,k_{a};p^{2},m^{2})\Upsilon
^{(n-a)}(k_{a+1},..,k_{n};m^{2})+  \nonumber \\
&&\sum_{a=1}^{n-1}\Upsilon ^{(n-a)}(k_{1},k_{2},..,k_{n-a};m^{2})\Theta
^{(a)}(k_{n-a+1},..,k_{n};p^{2},m^{2})+  \nonumber
\end{eqnarray}
\[
\sum_{a,b=1}^{n-2}\Upsilon ^{(b)}(k_{1},k_{2},..,k_{b};m^{2})\Theta
^{(a)}(k_{b+1},..,k_{a};p^{2},m^{2})\Upsilon
^{(n-a-b)}(k_{a+1},..,k_{n};m^{2})\Bigg\}
\]
From the above equation it becomes clear that the renormalization of the
self energy to $n^{th}$-order requires all finite functions defined in
previous self energy diagrams (up to $(n-1)^{th}$-order) as well as all the
divergent contributions of the three-point functions also to the $(n-1)^{th}$%
-order. We may associate a graphical representation to the equation above
and, in this way, compare with the BPHZ results. The first term in the
equation (\ref{145}) contains a sum of $n-1$ terms comprising $a$ finite
functions of the type $\Theta $ multiplied by the $n-a$ divergent
vertex-type functions. The second term is the symmetric to the first one
(the vertex functions and functions $\Theta $ swap sides). Finally the last
term contains vertex corrections to the left and to the right and finite
functions in the middle. This can be best visualized in the graph which
follows (figure 10).

Notice that in the present procedure no special treatment has been given to
the overlapping divergencies or to the nested ones, both appearing in the
self-energy. The reason is that the algebraic procedure produces only
disjoint divergent contributions.

In order to complete the renormalization of this theory we will still
consider two cases, both belonging to the second class defined previously.
Firstly we consider a specific case where two point functions explicitly
appear as subdivergences (see figure 11) and the other is an amplitude
containing an overlapping divergence diagram as substructure (figure 12). As
we mentioned before, the total integral contains the two point function
substructures in factorized form. We therefore effect the renormalization of
the internal propagators directly using the counterterms of order. In this
way we immediately obtain $\bar{\Gamma}$ . Let us first consider the case in
figure 11. This diagram contains subdiagrams involving nested two point
functions. Following the prescription which display the renormalized
contributions of previous orders we get 
\begin{eqnarray}
\bar{\Gamma} &=&i\bar{\Sigma}_{2}^{(n)}(p^{2}) \\
&=&(-i\lambda )^{2}(i)^{3}\int_{\Lambda }\frac{d^{6}k_{n}}{(2\pi )^{6}}\frac{%
1}{(k_{n}^{2}-m^{2})^{2}[(p-k_{n})^{2}-m^{2}]}  \nonumber \\
&&\times i\left\{ T_{k_{n}}^{2}\bar{\Sigma}_{2}^{(n-1)}(k_{n}^{2})+\delta
_{2}^{(n-1)}m^{2}-A_{2}^{(n-1)}k_{n}^{2}\right\}  \nonumber
\end{eqnarray}
with 
\begin{eqnarray}
i\bar{\Sigma}_{2}^{(n-1)}(k_{n}^{2}) &=&(-i\lambda )^{2}(i)^{3} \\
&&\int_{\Lambda }\frac{d^{6}k_{n-1}}{(2\pi )^{6}}\frac{1}{%
(k_{n-1}^{2}-m^{2})^{2}[(k_{n}-k_{n-1})^{2}-m^{2}]}  \nonumber \\
&&\times i\left\{ T_{k_{n-1}}^{2}\bar{\Sigma}_{2}^{(n-2)}(k_{n-1}^{2})+%
\delta _{2}^{(n-2)}m^{2}-A_{2}^{(n-2)}k_{n-1}^{2}\right\}  \nonumber
\end{eqnarray}
\begin{eqnarray*}
&&\bullet \\
&&\bullet \\
&&\bullet
\end{eqnarray*}
\begin{eqnarray}
i\bar{\Sigma}_{2}^{(2)}(k_{3}^{2}) &=&(-i\lambda )^{2}(i)^{3}\int_{\Lambda }%
\frac{d^{6}k_{2}}{(2\pi )^{6}}\frac{1}{%
(k_{2}^{2}-m^{2})^{2}[(k_{3}-k_{2})^{2}-m^{2}]} \\
&&\times i\left\{ T_{k_{2}}^{2}\Sigma ^{(1)}(k_{2}^{2})+\delta
^{(1)}m^{2}-A^{(1)}k_{2}^{2}\right\}  \nonumber
\end{eqnarray}
\begin{equation}
i\Sigma ^{(1)}(k_{2}^{2})=(-i\lambda )^{2}(i)^{2}/2\int_{\Lambda }\frac{%
d^{6}k_{1}}{(2\pi )^{6}}\frac{1}{(k_{1}^{2}-m^{2})[(k_{2}-k_{1})^{2}-m^{2}]}%
\cdot
\end{equation}
We can substitute the terms in brackets by renormalized function 
\begin{eqnarray}
\bar{\Sigma}_{2R}^{(n-1)}(k_{n}^{2}) &=&\Gamma _{R}^{(n-1)} \\
&=&T_{k_{n}}^{2}\bar{\Sigma}_{2}^{(n-1)}(k_{n}^{2})+\delta
_{2}^{(n-1)}m^{2}-A_{2}^{(n-1)}k_{n}^{2}\cdot  \nonumber
\end{eqnarray}
In this example the two point functions counterterms can be obtained for any
order $n>1$ as 
\begin{equation}
i\delta _{2}^{(n)}m^{2}=(-i\lambda )^{2}(i)^{3}\int_{\Lambda }\frac{%
d^{6}k_{n}}{(2\pi )^{6}}\frac{1}{(k_{n}^{2}-m^{2})^{3}}\left\{ i\bar{\Sigma}%
_{2R}^{(n-1)}(k_{n}^{2})\right\}
\end{equation}
\begin{eqnarray}
iA_{2}^{(n)} &=&(-i\lambda )^{2}(i)^{3}\{ \\
&&\frac{4}{6}g^{\mu \nu }\int_{\Lambda }\frac{d^{6}k_{n}}{(2\pi )^{6}}\frac{%
k_{\mu }k_{\nu }}{(k_{n}^{2}-m^{2})^{5}}\left\{ i\bar{\Sigma}%
_{2R}^{(n-1)}(k_{n}^{2})\right\}  \nonumber \\
&&-\int_{\Lambda }\frac{d^{6}k_{n}}{(2\pi )^{6}}\frac{1}{%
(k_{n}^{2}-m^{2})^{4}}\left\{ i\bar{\Sigma}_{2R}^{(n-1)}(k_{n}^{2})\right\}
\}\cdot  \nonumber
\end{eqnarray}

Finally we will consider type 3 diagram (which contain a type 1 diagram)
show in figure 12. \ The corresponding amplitude reads 
\begin{equation}
i\Sigma _{3}^{(n)}(p^{2})=(-i\lambda )^{2}(i)^{3}\int_{\Lambda }\frac{%
d^{6}k_{n}}{(2\pi )^{6}}\frac{1}{(k_{n}^{2}-m^{2})^{2}[(p-k_{n})^{2}-m^{2}]}%
\left\{ i\Sigma _{1}^{(n-1)}(k_{n}^{2})\right\} \cdot
\end{equation}
Note that since the structure $\Sigma _{1}^{(n-1)}(k_{n}^{2})$ can be
renormalized at $(n-1)^{th}$ order (see first example of order n), the $%
n^{th}$ order structure will also be renormalized by a $\Gamma
_{local}^{div} $. Its counterterms have the same form as in the previous
example. They are 
\begin{equation}
i\delta _{3}^{(n)}m^{2}=(-i\lambda )^{2}(i)^{3}\int_{\Lambda }\frac{%
d^{6}k_{n}}{(2\pi )^{6}}\frac{1}{(k_{n}^{2}-m^{2})^{3}}\left\{ i\Sigma
_{1R}^{(n-1)}(k_{n}^{2})\right\}
\end{equation}
\begin{eqnarray}
iA_{3}^{(n)} &=&(-i\lambda )^{2}(i)^{3}\{ \\
&&\frac{4}{6}g^{\mu \nu }\int_{\Lambda }\frac{d^{6}k_{n}}{(2\pi )^{6}}\frac{%
k_{\mu }k_{\nu }}{(k_{n}^{2}-m^{2})^{5}}\left\{ i\Sigma
_{1R}^{(n-1)}(k_{n}^{2})\right\}  \nonumber \\
&&-\int_{\Lambda }\frac{d^{6}k_{n}}{(2\pi )^{6}}\frac{1}{%
(k_{n}^{2}-m^{2})^{4}}\left\{ i\Sigma _{1R}^{(n-1)}(k_{n}^{2})\right\}
\}\cdot  \nonumber
\end{eqnarray}
Note that in this case the three point functions subdiagrams have been
renormalized together with the two point subdiagram, since it is contained
in the latter.

\section{Momentum routing independence}

In the examples of the previous sections we have chosen the momentum routing
in such way as to obtain the simplest form for the final expressions. Of
course, the counterterms so obtained must be independent of the particular
routing one chooses. In order to exemplify this we consider the last example
given (type 3 diagram). One of the possible choices for the momentum routing
would be to arrange the labels in such a way that external momentum is
present in an internal line of the diagram as follows 
\begin{eqnarray}
i\Sigma _{3}^{(n)}(p^{2}) &=&(-i\lambda )^{2}(i)^{3} \\
&&\int_{\Lambda }\frac{d^{6}k_{n}}{(2\pi )^{6}}\frac{1}{%
(k_{n}^{2}-m^{2})[(p-k_{n})^{2}-m^{2}]^{2}}\left\{ i\Sigma
_{1}^{(n-1)}((p-k_{n})^{2})\right\}  \nonumber
\end{eqnarray}
and also, 
\begin{equation}
i\Sigma _{3}^{(n)}(p^{2})=(-i\lambda )^{2}(i)^{3}\int_{\Lambda }\frac{%
d^{6}k_{n}}{(2\pi )^{6}}\frac{1}{(k_{n}^{2}-m^{2})^{2}[(p-k_{n})^{2}-m^{2}]}%
\left\{ i\Sigma _{1}^{(n-1)}(k_{n}^{2})\right\} \cdot
\end{equation}
These two labels must be equivalent, so that the amplitude is momentum
routing independent as it should. Note that if the amplitude were finite,
this could immediately be accomplished through a shift $p-k_{n}=k_{n}^{%
\prime }$. However, since the amplitude is quadratically divergent, shifts
are not allowed without the inclusion of surface terms. This point has been
extensively discussed in our method (see refs. \cite{13}\cite{14} ) and a
similar procedure can cure this problem in the present model. More difficult
would be theories with gauge symmetries and work along this line is in
progress. Note that in Dimensional Regularization the problem does not
appear since shifts are always allowed.

\section{Conclusion}

We have considered (in the self-energy) all possible complications which
usually appear in renormalization procedures: overlapping divergences,
nested divergencies and disjoint ones, all in the same graph at $n$-loops.
We have explicitly shown how these problems can be systematically resolved
order by order within our technique in a simple example, devoid of
symmetries abelian or not. However, several generals aspects of the method
can be learned already from this simple case :

there will always be a divergent (local) order dependent contribution. Also,
there will always be a finite contribution composed by the product of all
finite parts of $f_{j}$%
\'{}%
s. These two structures (divergent and finite) are typical of the $n^{th}$
order and poses no problem for renormalization.

As we have seen in the examples given, the identities we use in the
integrand leave us then with crossed products of divergent and finite
contributions. All possible combinations will appear and all of them can
either be recognized as structures (finite or divergent) already encountered
in lower order amplitudes or they will give a finite contribution.

In summary we have presented a completely algebraic method of perturbative
renormalization. All counterterms appear automatically, and presents a close
correspondence to BPHZ where graphical representations are essential. We
feel that if our method also works for abelian and non abelian theories this
could be a major advance. At the one loop level we can preserve gauge
symmetry if use is made of relations involving divergent integrals of the
same degree of divergence \cite{14},\cite{13}. The difference among those
integrals is source of both ambiguities and symmetry violations .The results
we got at one loop are encouraging We are working on the application of this
method at the two loops level the Quantum Electrodynamic (QED).

We see that the application of this method leads to a relatively simple
renormalization procedure. There is no need for a graphic representation of
the relevant contributions. When a diagram has divergent subdiagrams the
subdivergences need not be previously identified because they will appear in
an algebraic way. The procedure display all relevant subdivergences.
Investigations of advantages of the present method (if at all!) over the
existing ones remain as our main research interest right now.

\section*{Acknowledgments}

The work of \ M.C. Nemes\ was partially supported by CNPq, FAPEMIG. The work
of \ S.R. Gobira was partially supported by FAPEMIG.

\begin{center}
\newpage {\em FIGURE CAPTIONS}
\end{center}

\begin{itemize}
\item  Figure 1: One-loop vertex correction $-iV^{(1)}(p,p^{\prime })$

\item  Figure 2:\ The two-loop vertex correction contribution $%
-iV_{1}^{(2)}(p,p^{\prime })$

\item  Figure 3:\ The two-loop vertex correction contribution $%
-iV_{2}^{(2)}(p,p^{\prime })$

\item  Figure 4:\ The two-loop vertex correction contribution $%
-iV_{3}^{(2)}(p,p^{\prime })$

\item  Figure 5:\ The n-loop vertex correction contribution $%
-iV_{1}^{(n)}(p,p^{\prime })$

\item  Figure 6: One-loop self energy $i\Sigma ^{(1)}(p^{2})$

\item  Figure 7: The two-loop self energy contribution $i\Sigma
_{1}^{(2)}(p^{2})$

\item  Figure 8: The two-loop self energy contribution $i\Sigma
_{2}^{(2)}(p^{2})$

\item  Figure 9: The n-loop self energy contribution $i\Sigma
_{1}^{(n)}(p^{2})$

\item  Figure 10: Graphic representation of equation (145).

\item  Figure 11: The n-loop self energy contribution $i\Sigma
_{2}^{(n)}(p^{2})$

\item  Figure 12: The n-loop self energy contribution $i\Sigma
_{3}^{(n)}(p^{2})$

\newpage 
\begin{figure}[tbp]
\centering
\includegraphics[scale=1.0]{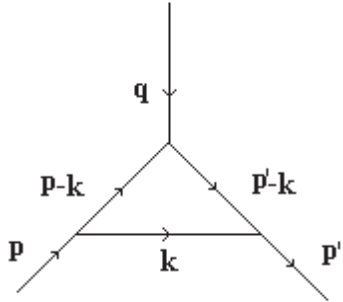}
\caption{One-loop vertex correction $-iV^{(1)}(p,p^{\prime })$}
\end{figure}
\end{itemize}

\begin{figure}[]
\centering
\includegraphics[scale=1.0]{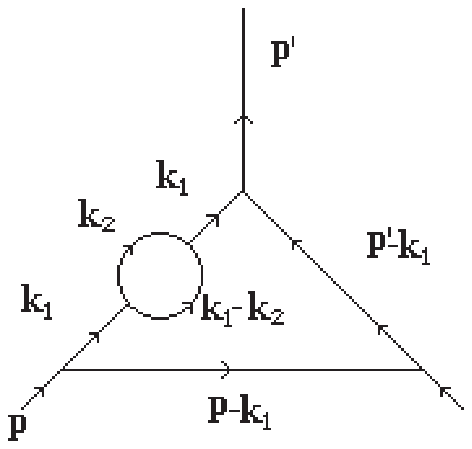}
\caption{The two-loop vertex correction contribution $-iV_{1}^{(2)}(p,p^{%
\prime })$}
\end{figure}

\begin{figure}[]
\centering
\includegraphics[scale=1.0]{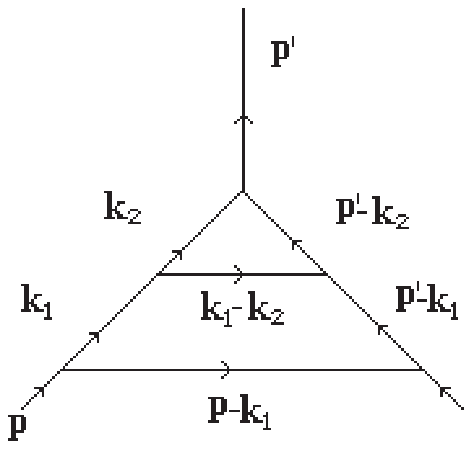}
\caption{The two-loop vertex correction contribution $-iV_{2}^{(2)}(p,p^{%
\prime })$}
\end{figure}

\begin{figure}[]
\centering
\includegraphics[scale=1.0]{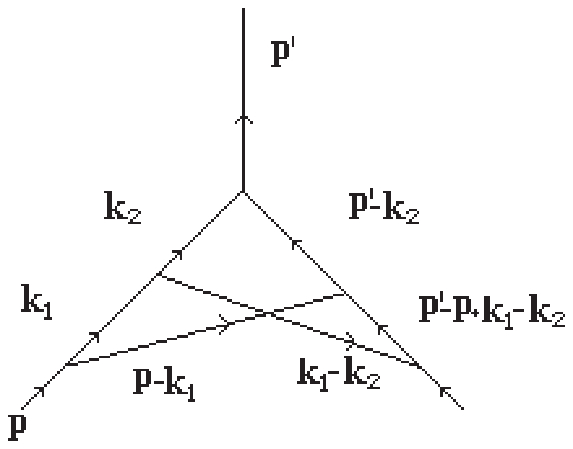}
\caption{The two-loop vertex correction contribution $-iV_{3}^{(2)}(p,p^{%
\prime })$}
\end{figure}

\begin{figure}[]
\centering
\includegraphics[scale=1.0]{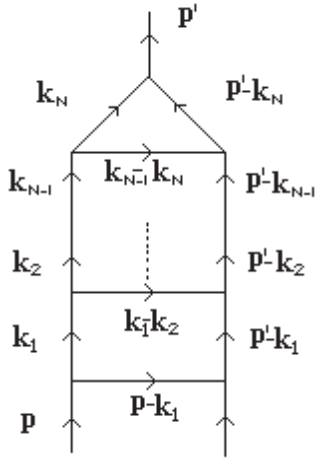}
\caption{The n-loop vertex correction contribution $-iV_{1}^{(n)}(p,p^{%
\prime })$}
\end{figure}

\begin{figure}[]
\centering
\includegraphics[scale=1.0]{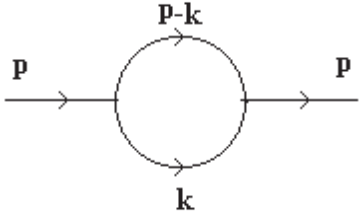}
\caption{One-loop self energy $i\Sigma ^{(1)}(p^{2})$}
\end{figure}

\begin{figure}[]
\centering
\includegraphics[scale=1.0]{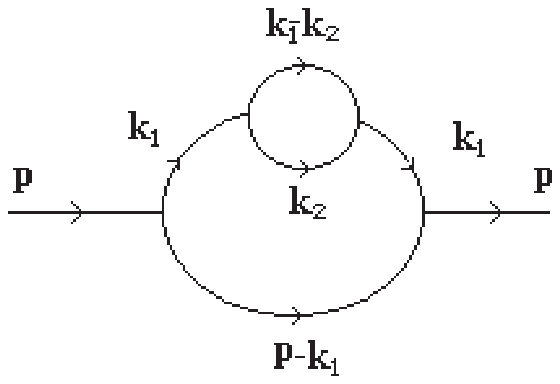}
\caption{The two-loop self energy contribution $i\Sigma _{1}^{(2)}(p^{2})$}
\end{figure}

\begin{figure}[]
\centering
\includegraphics[scale=1.0]{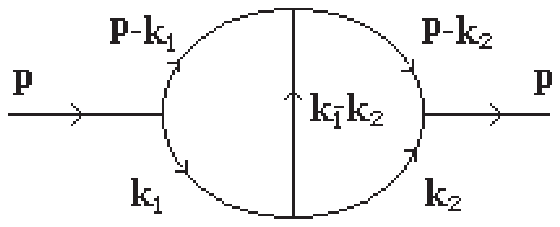}
\caption{The two-loop self energy contribution $i\Sigma _{2}^{(2)}(p^{2})$}
\end{figure}

\begin{figure}[]
\centering
\includegraphics[scale=1.0]{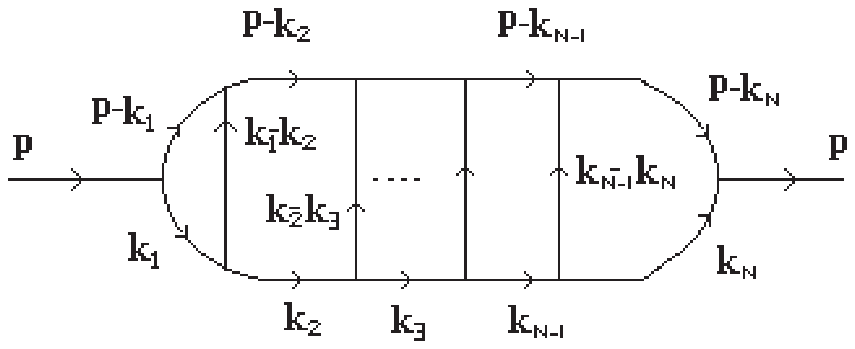}
\caption{The n-loop self energy contribution $i\Sigma _{1}^{(n)}(p^{2})$}
\end{figure}

\begin{figure}[]
\centering
\includegraphics[scale=0.9]{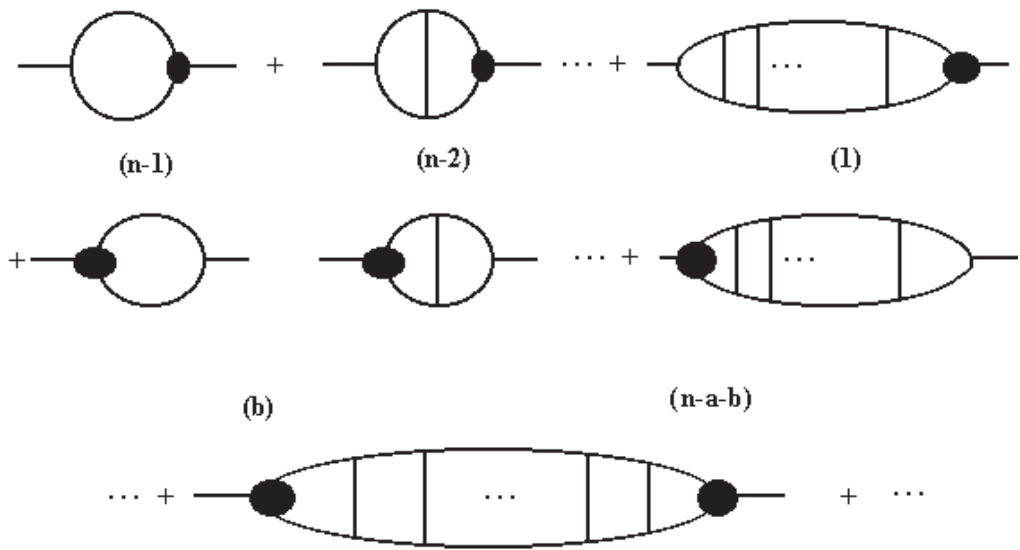}
\caption{Graphic representation of equation (145)}
\end{figure}

\begin{figure}[]
\centering
\includegraphics[scale=0.9]{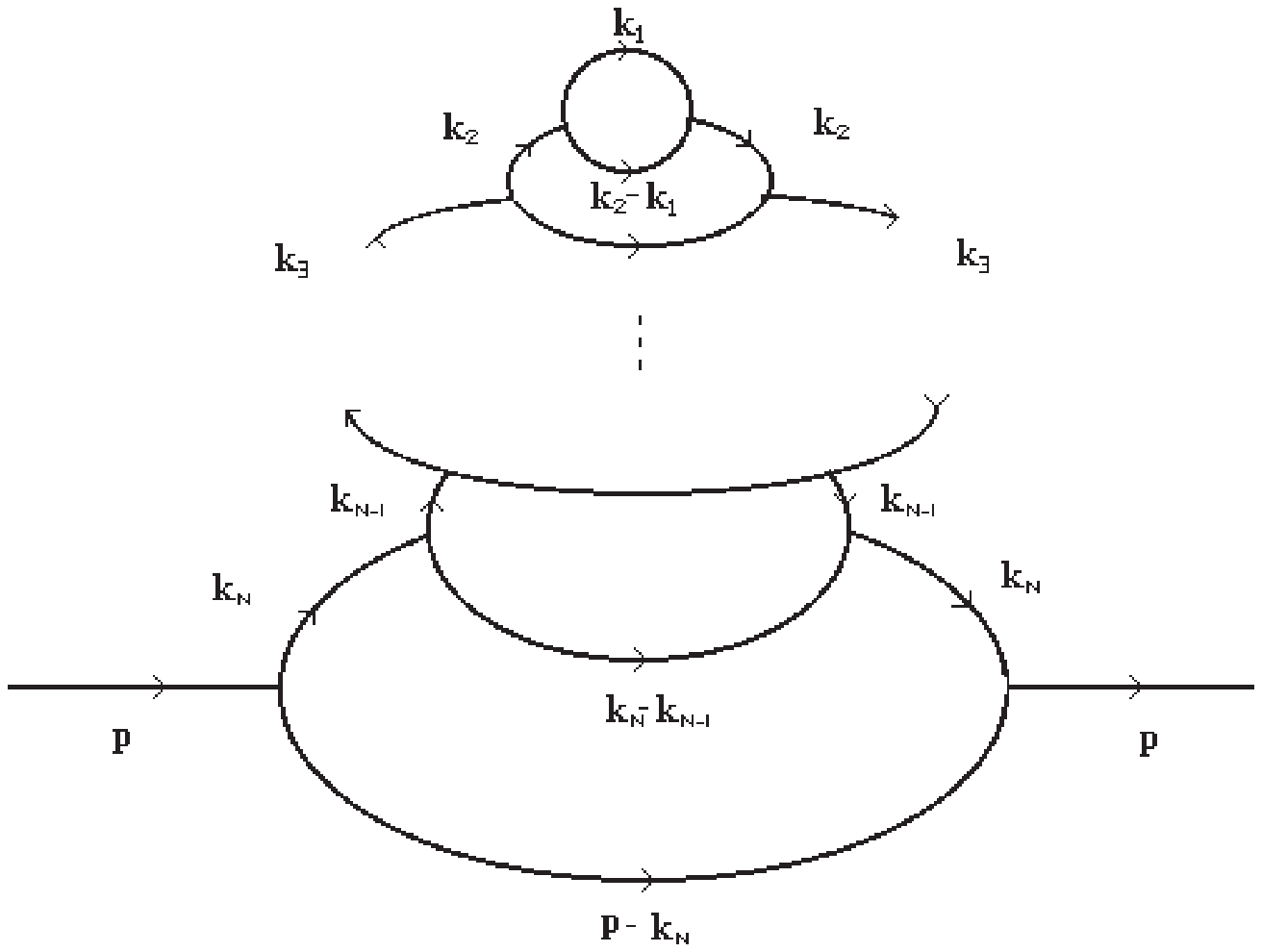}
\caption{The n-loop self energy contribution $i\Sigma _{2}^{(n)}(p^{2})$}
\end{figure}

\begin{figure}[tbp]
\centering
\includegraphics[scale=1.2]{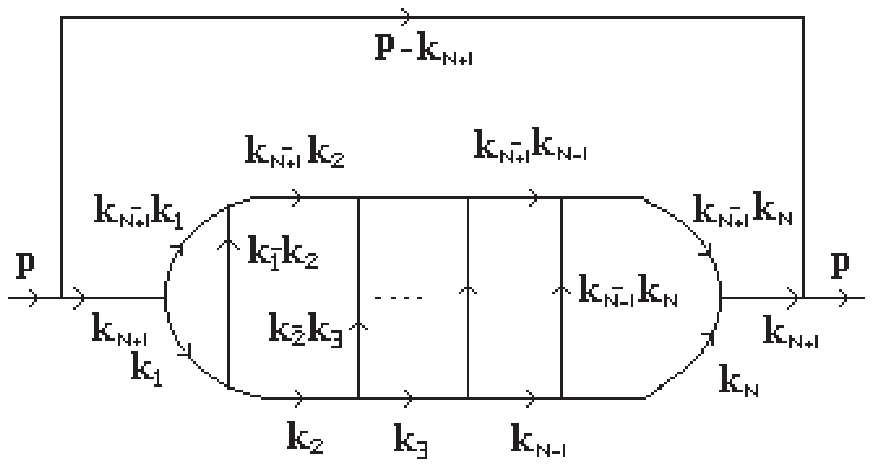}
\caption{The n-loop self energy contribution $i\Sigma _{3}^{(n)}(p^{2})$}
\end{figure}

\end{document}